\documentclass[a4paper,twoside]{article}

\usepackage{epsfig}
\usepackage{subcaption}
\usepackage{calc}
\usepackage{amssymb}
\usepackage{amstext}
\usepackage{amsmath}
\usepackage{amsthm}
\usepackage{multicol}
\usepackage{pslatex}
\usepackage{apalike}
\usepackage{SCITEPRESS}     
\usepackage{amsmath,amssymb,amsbsy,amsfonts,amsthm,latexsym,
            amsopn,amstext,amsxtra,euscript,amscd,color,breqn,mathrsfs}
            
\newtheorem{thm}{Theorem}
\newtheorem{lem}[thm]{Lemma}

\newtheorem{defn}[thm]{Definition}

\newtheorem{remark}[thm]{Remark}
\newtheorem{example}[thm]{Example}
      
\newcommand{\Tr}{{\rm Tr}}
\newcommand{\Trn}{{\rm Tr}_n}

\newcommand{\Trm}{{\rm Tr}_m}
\def\cB{{\mathcal B}}

\def\C{{\mathbb C}}
\def\F{{\mathbb F}}
\def\cB{{\mathcal B}}
\def\cC{{\mathcal C}}

\def\cW{{\mathcal W}}

\newcommand{\vwht}[3]{\mathcal{W}_{#1}(#2,#3)}

\def\ee{{\bf e}}

\def\xx{{\mathbf x}}

\begin{document}

\title{An extension of the avalanche criterion in the context of $c$-differentials}

\author{\authorname{\Large P\aa l Ellingsen\sup{1}, Constanza Riera\sup{1}, Pantelimon~St\u anic\u a\sup{2}, \Large Anton Tkachenko\sup{1}}
\affiliation{\sup{1}Department of Computer Science,
 Electrical Engineering and Mathematical Sciences,
   Western Norway University of Applied Sciences,
  5020 Bergen, Norway}
\affiliation{\sup{2}Department of Applied Mathematics,
Naval Postgraduate School,
Monterey, CA 93943--5216,  USA}
\email{\{pel, csr, atk\}@hvl.no, pstanica@nps.edu}
} 

\abstract{The Strict Avalanche Criterion (SAC) is a property of vectorial Boolean functions that is used in the construction of strong S-boxes.
We show in this paper how to generalize the concept of SAC to address possible $c$-differential attacks, in the realm of finite fields. We define the concepts of $c$-Strict Avalanche Criterion ($c$-SAC) and $c$-Strict Avalanche Criterion of order $m$ ($c$-SAC($m$)), and generalize results of \cite{LC07}. We also show computationally how the new definition is not equivalent to the existing concepts of $c$-bent$_1$-ness \cite{SGGRT20}, nor (for $n=m$) PcN-ness~\cite{EFRST20} }

\onecolumn \maketitle \normalsize \setcounter{footnote}{0} \vfill

\section{\uppercase{Introduction}}
\label{sec:introduction}

A Substitution-box (S-Box) is one of the most important elements that are used to provide attack resistance to block encryption algorithms. 
An S-box performs substitution on its input symbols, and together with permutations, they are typically used to obscure the relationship between the plaintext, the key and the ciphertext of an encryption algorithm.
The attack resistance of an S-box depends on many different factors, but one of the more important ones is the ability of the S-box to induce a significant change in the output of the box from a small change in the input. This is called the avalanche effect. In general, if an S-box does not have this avalanche effect, it will result in a lack of randomization in the algorithm that may be used as part of an attack on the algorithm. A primary strategy for attacking cryptographic algorithms with weak randomization properties are the so-called differential attacks \cite{biham1991differential}, \cite{biham2012differential}. The {\em Strict Avalanche Criterion} ({\em SAC}) is a more refined property of S-boxes derived from the general avalanche property. SAC was introduced in \cite{WT85} in the context of S-boxes, described by vectorial Boolean functions, as follows: a vectorial Boolean function satisfies SAC if and only if whenever a single input bit of a coordinate is complemented, each of its output bits changes with probability $1/2$; i.e. given $F:\F_2^n\rightarrow \F_2^m$ ($\F_2$ is the two-element field and $\F_2^k$ is a vector space of dimension $k$ over $\F_2$), the function $F=(F_1,\ldots,F_m)$ satisfies SAC if and only if the probability $Prob(F_i(\xx\oplus\ee_i)\oplus F_i(\xx)=1)=\frac{1}{2},\ \forall i=1,\ldots,m$, where $\ee_i$ is the standard basis vector with 1 in component $i$ and 0 in all other components. 

In the paper \cite{EFRST20}, we defined the concept of $c$-differential uniformity, that may leave ciphers vulnerable to differential cryptanalysis. This concept has also been explored further for power functions with good properties for S-box design in \cite{SPR21}, \cite{SG20}, \cite{yan2020power}, to cite only a few papers among the many that appeared in a short time on the topic.
In this paper, we extend the Strict Avalanche Criterion to address new attacks that might stem from such use of the $c$-differential.


Surely, the { Strict Avalanche Criterion} can be defined for (vectorial or single output) Boolean and $p$-ary functions. Throughout this paper, we will take the primitive root of unity, $\zeta=\zeta_p= e^{\frac{2\pi i}{p}}$, for any prime~$p$.


\begin{defn}\textup{\cite{LC07}}
\label{defnvectors} 
Let $wt(a)$, for $a\in\F_p^n$ be the Hamming weight of $a$, that is, the number of nonzero components of $a$. Then, \begin{itemize}
\item $f:\F_p^n\rightarrow \F_p$ fulfills the {\em Strict Avalanche Criterion (SAC)} if and only if $Prob(f(x+a)-f(x)=b)=\frac{1}{p},\ \forall a\in\F_p^n,b\in\F_p,\,wt(a)=1$. Equivalently, $f$ fulfills SAC if and only if 
\begin{equation}
\sum_{x\in\F_p^n}\zeta^{f(x+a)-f(x)}=0,\ \forall a\in\F_p^n,\,wt(a)=1.
\end{equation} 

\item For vectorial $p$-ary functions, this is defined componentwise: a vectorial $p$-ary function $F=(F_0,\ldots,F_{m-1}):\F_p^n\rightarrow \F_p^m$ fulfills the {\em Strict Avalanche Criterion  } if and only if $Prob(F_i(x+a)-F_i(x)=b)=\frac{1}{p},\ \forall i=0,\ldots,m-1, \forall a\in\F_p^n,b\in\F_p,\,wt(a)=1$. Equivalently, $F$ fulfills SAC if and only if $, \forall i=0,\ldots,m-1,$
$$\sum_{x\in\F_p^n}\zeta^{F_i(x+a)-F_i(x)}=0,\ \forall a\in\F_p^n,\,wt(a)=1.$$

\end{itemize}
\end{defn}
In this paper, we present a new form of Strict Avalanche Criterion based on $c$-differentials (as defined in~\cite{EFRST20}), and extend the results of \cite{LC07} to this new criterion. We need first to rewrite the definition of SAC in the context of finite fields, since the new criterion is more naturally defined in that context.

Let $g$ be a generator of the finite field $\F_{p^k}$. For any $k$, we use the 
identification $M_g:\F_p^k\rightarrow \F_{p^k}$, defined as $M_g((x_0,\ldots,x_{k-1}))=x_0+x_1g+\cdots+x_{k-1}g^{k-1}$. Then, $wt(\alpha)=1$ if and only if $M_g(\alpha)=\alpha_tg^t$ for some $t=0,\ldots,k-1$, $\alpha_t\in\F_p^*$. The components of a vectorial $p$-ary function $F:\F_{p^n}\rightarrow \F_{p^m}$ are $\Tr_m(bF(x))$, where $\Tr_m:\F_{p^m}\to \F_p$ is the absolute trace function, given by $\displaystyle \Tr_m(x)=\sum_{i=0}^{m-1} x^{p^i}$ (we will denote it by $\Tr$, if the dimension is clear from the context). So, it is natural to define the Strict Avalanche Criterion 
relating it to the derivative of~$F$\footnote{We have not found a definition for SAC in the finite fields context in the literature, but we do not claim that this is necessarily new.}:
\begin{defn}\label{defnfinitefields}
Let $F:\F_{p^n}\rightarrow \F_{p^m}$ a $p$-ary $(n,m)$-function. Let $g$ be a generator of $\F_{p^n}$. We say that $F$ fulfills  SAC if and only if $$\sum_{x\in\F_p^n}\zeta^{\Tr_m(b(F(x+a)-F(x)))}=0, \mbox{ for  all }    b\in\F_{p^m}^*,\,a=a_tg^t,$$ for some $t=0,\ldots,k-1$, $a_t\in\F_p^*$.
\end{defn}

NB: Given a $p$-ary $(n,m)$-function   $F:\F_{p^n}\to \F_{p^m}$,  the derivative of $F$ with respect to~$a \in \F_{p^n}$ is the  function
\[
 D_{a}F(x) =  F(x + a)- F(x), \mbox{ for  all }  x \in \F_{p^n}.
\] Using this notation, fulfills  SAC if and only if $$\sum_{x\in\F_p^n}\zeta^{\Tr_m(bD_aF(x))}=0, \mbox{ for  all }   b\in\F_{p^m}^*,\,a=a_tg^t\in\F_{p^n}^*,$$ for some $t=0,\ldots,k-1$, $a_t\in\F_p^*$. 

\begin{remark} Note that Definition~\textup{\ref{defnfinitefields}} is more restrictive than Definition~\textup{\ref{defnvectors}}. The reason for this is that, while in the usual definition (Definition~\textup{\ref{defnvectors}}) the output components are considered independently, in Definition~\textup{\ref{defnfinitefields}} the condition is for the derivative as a single object; in fact, for $n=m$, the condition of Definition~\textup{\ref{defnfinitefields}} is equivalent to all derivatives $D_a F(x)$ (for $a=a_tg^t$) being permutation polynomials (see~\textup[Theorem 7.7]{\cite{LN97}}), and, in fact, as Lemma~\textup{\ref{balanced}} shows (taking $c=1$), Definition~\textup{\ref{defnfinitefields}} is equivalent to the balancedness of the derivative itself. For example, the function $F:\F_2^2\rightarrow\F_2^2$ defined by $F(x_0,x_1)=(x_0x_1,x_0x_1)$ fulfills SAC according to Definition~\textup{\ref{defnvectors}}, since, for each component, the derivatives with respect to  $a=(0,1)$ and $a=(1,0)$ are balanced. However, if we map the function $F:\F_2^2\rightarrow\F_2^2$ to the function $F':\F_4\rightarrow\F_4$ by applying the map $M_g$ to its input and output we see that $F'$ has values $F'(0)=0,F'(g)=0,F'(1)=0,F'(g^2)=g^2$. It is easy to see that neither derivative $D_1F'(x)$ nor $D_gF'(x)$ are permutation polynomials. Thus, $F'$ does not fulfill SAC under Definition~\textup{\ref{defnfinitefields}}. Furthermore, while there exist functions $F:\F_2^n\rightarrow \F_2^n$ that fulfill SAC under Definition~\textup{\ref{defnvectors}} (at any rate, for even dimension), there exists no function $F':\F_{2^n}\rightarrow\F_{2^n}$ that fulfills SAC under Definition~\textup{\ref{defnfinitefields}}, since $F'(x+a)-F'(x)$ has the same values for $x$ and $x+a$, and can therefore never be a permutation. However, it is not an empty definition, if $m\neq n$, as we see below. 
\end{remark}
\begin{example} The function $F:\F_{4}\rightarrow\F_{2}$ defined by its values $F(0)=0,F(g)=0,F(1)=0,F(g^2)=1$  fulfills SAC under Definition~\textup{\ref{defnvectors}}, since here $b=1$, $a=1$ or $a=g$, and 
$$\sum_{x\in\F_4}(-1)^{\Tr_1(bD_1F(x))}=\sum_{x\in\F_4}(-1)^{D_1F(x)}=0$$ and
$$\sum_{x\in\F_4}(-1)^{\Tr_1(bD_gF(x))}=\sum_{x\in\F_4}(-1)^{D_gF(x)}=0.$$
\end{example}
We recall below the differential extension from~\cite{EFRST20}, in the context of  finite fields.
\begin{defn}~\textup{\cite{EFRST20}}
Given a $p$-ary $(n,m)$-function   $F:\F_{p^n}\to \F_{p^m}$, and $c\in\F_{p^m}$, the (multiplicative) $c$-derivative of $F$ with respect to~$a \in \F_{p^n}$ is the  function
\begin{equation}
 _cD_{a}F(x) =  F(x + a)- cF(x), \mbox{ for  all }  x \in \F_{p^n}.
\end{equation}
(Note that, if   $c=1$, then we obtain the usual derivative, and, if $c=0$ or $a=0$, then we obtain a shift (input, respectively, output) of the function.)
\end{defn}

It is natural to consider then an extension of the Strict Avalanche Criterion (SAC) using this new derivative.

\section{\uppercase{The $c$-Strict Avalanche Criterion ($c$-SAC)}}
\label{sec:The $c$-Strict Avalanche Criterion ($c$-SAC)}
In this section, we extend the Strict Avalanche Criterion (SAC) to address new attacks that might stem from the use of the $c$-differential.

\begin{defn} Let $F:\F_{p^n}\rightarrow \F_{p^m}$ be a $p$-ary $(n,m)$-function. We say that $F$ fulfills the $c$-Strict Avalanche Criterion ($c$-SAC) if and only if  $\sum_{x\in\F_p^n}\zeta^{\Tr_m(b(F(x+a)-cF(x)))}=\sum_{x\in\F_p^n}\zeta^{\Tr_m(bF(x+a))-\Tr_m(cbF(x))}=0$ for all $b\in\F_{p^m}^*,\,a=a_tg^t\in\F_{p^n}^*,$ for some $t=0,\ldots,k-1$, $a_t\in\F_p^*$.

\end{defn}

In \cite{SGGRT20}, for $F\in\cB_{n,p}^m$ (the set of all functions  from $\F_{p^n}\rightarrow \F_{p^m}$) and fixed $c\in\F_{2^m} $, we define the {\em $c$-crosscorrelation} at $u\in\F_{p^n}, b\in\F_{p^m}$ by 
\begin{equation}
{_c}\cC_{F,G}(u,b)=\sum_{x \in \F_{p^n}} \zeta_{p}^{\Trm(b(F(x+u)  - c G(x)))}
\end{equation}
and the corresponding {\em $c$-autocorrelation} at $u\in\F_{p^n}$, ${_c}\cC_{F}(u,b)={_c}\cC_{F,F}(u,b)$. 

Using this, we say that for $F\in\cB_{n,p}^m$ and fixed $c\in\F_{2^m} $, $F$ fulfills  the $c$-Strict Avalanche Criterion ($c$-SAC) if and only if ${_c}\cC_{F}(a,b)=0,\ \forall b\in\F_{p^m}^*,\,a=a_tg^t\in\F_{p^n}^*$ for some $t=0,\ldots,k-1$, $a_t\in\F_p^*$. 

\begin{remark} Note that, for $n=m$, the Perfect $c$-Nonlinear (PcN) class defined in~\textup{\cite{EFRST20}} is a subclass of the set of functions fulfilling $c$-SAC, and, in general, its generalization, the $c$-bent$_1$ class, defined in~\textup{\cite{SGGRT20}}, is a subclass of $c$-SAC. However, as we show in Section~\textup{\ref{computations}}, these subclasses are strict, and we can find examples of $(n,n)$-vectorial $p$-ary functions that fulfill $c$-SAC for some $c$ but are not PcN (which, for $n=m$, is equivalent to $c$-bent$_1$) for that value of $c$, for both even and odd characteristics.
\end{remark}

\section{\uppercase{Theoretical results on  the $c$-Strict Avalanche Criterion}}
\label{sec:Theoretical results on  the $c$-Strict Avalanche Criterion}

Note that, as in the classical case, the correlation condition and the balancedness are equivalent\footnote{Note that, as stated before, for the case $n=m$, this result is given in~\cite[Theorem 7.7]{LN97}.}:

\begin{lem} \label{balanced} Let $F:\F_{p^n}\rightarrow \F_{p^m}$ a $p$-ary $(n,m)$-function. Then, $F$ fulfills $c$-SAC if and only if all the traces of multiples of $c$-differentials with respect to any $a$ of $p$-ary weight $1$ are balanced, i.e.   $\Trm( _cD_{a}bF(x)) $ is balanced, for all $b\in\F_{p^m}^*,\,a=a_tg^t\in\F_{p^n}^*$, for some $t=0,\ldots,k-1$, $a_t\in\F_p^*$.
\end{lem}
\begin{proof} We follow the proof of \cite[Theorem 2.5]{SGGRT20}, and include it here for the convenience of the reader.

 With $c\in\F_{p^n}$ constant, for every $u\in\F_{p^n}, b\in\F_{p^m}$, $0\leq j\leq p-1$, we let  $S_{j,c}^{u,b}=\{x\in\F_{p^n}\,|\, \Trm(b(F(x+u)-cF(x)))=j\}$.  We will use below that the order of the cyclotomic polynomial of index $p^m$ is  $\phi(p^m)=p^{m-1}(p-1)$.
 
 First, recall that the $p^k$-cyclotomic polynomial is $\phi_{p^k}(x)=1+x^{p^{k-1}}+x^{2p^{k-1}}+\cdots+x^{(p-1)p^{k-1}}$. In particular, we deduce that $\zeta_p^{p-1}=-(1+\zeta_p+\cdots+\zeta_p^{p-2})$.
 If   $u\in\F_{p^n}^*$ such that $u=u_tg^t$ for some $t=0,\ldots,k-1$, $u_t\in\F_p^*$, $  b\in\F_{p^m}^*$, and $F$ fulfills $c$-SAC, then
 \allowdisplaybreaks

\begin{equation} 
\begin{split}
0=&{_c}\cC_F(u ,b)
 =\sum_{x \in \F_{p^n}} \zeta_{p}^{\Trm(b(F(x+u)  - c  F(x )))}\\
 =&\sum_{j=0}^{p-1} |S_{j,c}^{u,b}| \zeta_{p}^{j}
  =  \sum_{j=0}^{p-2}\left(|S_{j,c}^{u,b}|-  |S_{p-1,c}^{u,b}| \right) \zeta_{p}^{j}.
 \end{split}
\end{equation}

 The  extension $\mathbb{Q} \stackrel{p-1}{\hookrightarrow}\mathbb{Q}(\zeta_{p})$ has degree $p-1$ and the elements in following set $\left\{\zeta_{p}^j\,|\,0\leq j\leq p-2\right\}$  are   linearly independent in $ \mathbb{Q}(\zeta_{p})$ over  $\mathbb{Q}$, therefore the coefficients in the displayed expression are zero, that is, that for all $0\leq j\leq p-2$,
$ |S_{j,c}^{u,b}|=|S_{p-1,c}^{u,b}| $. Summarizing, for any $0\leq j\leq p-1$, the cardinality of the set $S_{j,c}^{u,b}$ is independent of~$j$, and so, for all $c,b,u\neq 0$ fixed, the function $x\mapsto \Trm(b(F(x+u)-cF(x))$ is balanced for all $u=u_tg^t$ for some $t=0,\ldots,k-1$, $u_t\in\F_p^*$, $  b\in\F_{p^m}^*$.

If $x\mapsto \Trm(b(F(x+u)-cF(x))$ is balanced, by reversing the argument, we find that $f$ fulfills $c$-SAC.
\end{proof}

This means that  $F$ fulfills $c$-SAC if and only if any of the following equivalent conditions are fulfilled:
\begin{enumerate}
\item ${_c}\cC_{F}(a,b)=0\ \forall b\in\F_{p^m}^*,\,a=a_tg^t\in\F_{p^n}^*$, for some $t=0,\ldots,k-1$, $a_t\in\F_p^*$.
\item the function $x\mapsto \Trm(b(F(x+a)-cF(x))$ is balanced, for all $ b\in\F_{p^m}^*,\,a=a_tg^t\in\F_{p^n}^*$, for some $t=0,\ldots,k-1$, $a_t\in\F_p^*$.
\end{enumerate}

 The (vectorial) Walsh transform $\vwht{F}{a}{b}$ of an $(n,m)$-function $F:\F_{p^n}\to \F_{p^m}$ at $a\in \F_p^n, b\in \F_p^m$ is the Walsh-Hadamard transform of its component function ${\rm Tr}_m(bF(x))$ at $a$, that is,
\begin{equation}
  \vwht{F}{a}{b}=\sum_{x\in\F_{p^n}} \zeta_p^{\Trm(bF(x))-\Trn(ax)}.
\end{equation}
We can extend Lemma 3.3 of \cite{LC07}, using \cite[Lemma 2.2]{SGGRT20} (the proof follows directly from \cite[Lemma 2.2]{SGGRT20} and it is omitted):
\begin{lem} We have $h(u,b)={_c}\cC_{F}(u,b )$ if and only if $\sum_{u  \in \F_{p^n}}h(u,b )\zeta_p^{-\Trn(ux)}=\cW_F(x,b)\overline{\cW_{F}(x,bc)}
 $.
\end{lem}
%
%
%
 
 This implies that:
 
\begin{lem} Let $F:\F_{p^n}\rightarrow \F_{p^m}$ a $p$-ary $(n,m)$-function. Then, $F$ satisfies $c$-SAC if and only if $\sum_{y\in \F_{p^n}}\cW_F(y,b)\,\overline{\cW_{F}(y,bc)}\,\zeta_p^{\Trn(ay)}=0,\ \forall b\in\F_{p^m}^*,\,a=a_tg^t\in\F_{p^n}^*$,  for some $t=0,\ldots,k-1$, $a_t\in\F_p^*$.
\end{lem}
\begin{proof} By \cite[Lemma 2.2]{SGGRT20}, $\sum_{y\in \F_{p^n}}\cW_F(y,b)\overline{\cW_{F}(y,bc)}\zeta_p^{\Trn(ay)}=\cC_{F}(a,b )$. The result follows.
\end{proof}

Let $U,T:\F_{p^n}\times\F_{p^n}\rightarrow \C$. We define   the left convolution by $(U\star T)(x,y)=\sum_{z\in \F_{p^n}}U(x-z,y)T(z,y)$. Let $F,G: \F_{p^n}\rightarrow \F_{p^m}$. Then, it is easy to show that $$
(\cW_F\star\cW_G)(a,b)=p^n\cW_{F+G}(a,b)
$$

and so, 
\begin{equation}
(\cW_F\star\cW_{-F})(a,b)=
\begin{cases}
p^{2n}&\text{ if } a=0\\
0&\text{ if }a\neq0.
\end{cases}
\end{equation}

We will show below  that the $c$-SAC is preserved by extended-affine (EA) equivalence, where EA-equivalence is defined as follows:

\begin{defn}\textup{\cite{CP19}} Two functions $F,G: \F_{p^n}\rightarrow\F_{p^m}$ are extended-affine equivalent (EA-equivalent) if and only if there exist $\alpha\in\F_{p^m}^*,\,e\in\F_{p^m},\,\beta\in\F_{p^n}^*,\,d\in\F_{p^n}$ such that $G(x)=\alpha F(\beta x+d)+e$.
\end{defn}

The next theorem is a generalization of Theorems 3.6 and  3.7 of \cite{LC07}.
\begin{thm} 
The $c$-SAC is preserved under the EA-equivalence.
\end{thm}
\begin{proof} 
We need to prove that  $F$ satisfies $c$-SAC if and only if $G(x)=\alpha F(\beta x+d)+e$ satisfies $c$-SAC, where $\alpha\in\F_{p^m}^*,\,e\in\F_{p^m},\,\beta\in\F_{p^n}^*,\,d\in\F_{p^n}$.
 Thus, $G$ satisfies $c$-SAC if and only if $\sum_{x\in\F_p^n}\zeta^{\Tr_m(b(G(x+a)-cG(x)))}=0,\ \forall b\in\F_{p^m}^*,\,a=a_tg^t\in\F_{p^n}^*$, for some $t=0,\ldots,k-1$, $a_t\in\F_p^*$. Calling $y=\beta x+d$, we have that 
\begin{equation}
\begin{split}
& \sum_{x\in\F_p^n}\zeta_p^{\Tr_m(b(G(x+a)-cG(x)))}=\\
& = \sum_{x\in\F_p^n}\zeta_p^{\Tr_m(b(\alpha F(\beta x+d+a)+e-c(\alpha F(\beta x+d)+e))}\\
& =\zeta_p^{\Tr_m(b(1-c)e)}\sum_{y\in\F_p^n}\zeta_p^{\Tr_m(b\alpha(F(y+a)-cF(y)))}.
\end{split}
\end{equation}
The theorem follows.
\end{proof}

For the next result, which generalizes Theorem 3.8 of \cite{LC07}, we need to introduce some notations. Let $n=n_1+n_2$, and $g,g_1,g_2$ be generators of $\F_{p^{n}},\F_{p^{n_1}},\F_{p^{n_2}}$, respectively. Then, we can write any element $z$ of $\F_{p^{n}}$ as $z=x_0+x_1g+\cdots+x_{n_1-1}g^{n_1-1}+y_0g^{n_1}+y_1g^{n_1+1}+\cdots+y_{n_2}g^{n_1+n_2-1}$. We define then $\sigma_1:\F_{p^{n}}\rightarrow\F_{p^{n_1}}$ as $\sigma_1(z)=x_0+x_1g_1+\cdots+x_{n_1-1}g_1^{n_1-1}$ and $\sigma_2:\F_{p^{n}}\rightarrow\F_{p^{n_2}}$ as $\sigma_2(z)=y_0+y_1g_2+\cdots+y_{n_2-1}g_2^{n_1-1}$. It is easy to see that $\sigma_1$ and $\sigma_2$ are linear over $\F_p$. 

\begin{thm} 
Let $F$ be an $(n_1,m)$-vectorial $p$-ary function, and $G$ be an $(n_2,m)$-vectorial $p$-ary function. We define an $(n,m)$-vectorial $p$-ary function by $H(z)=F(\sigma_1(z))+G(\sigma_2(z))$. Then, $H$ fulfills SAC  if and only if both $F$ and $G$ fulfill SAC.
\end{thm}
\begin{proof}
We write
\small
\begin{equation}
\begin{split}
&\sum_{z\in\F_p^n}\zeta_p^{\Tr_m(b(H(z+a)-cH(z)))}= \\
&\sum_{z\in\F_p^n}\zeta_p^{\Tr_m(b(F(\sigma_1(z+a))-cF(\sigma_1(z))))}\zeta_p^{\Tr_m(b(G(\sigma_2(z+a))-cF(\sigma_2(z))))}.
\end{split}
\end{equation}
\normalsize
Now, $\sigma_i(z+a)=\sigma_i(z)+\sigma_i(a)$. Since $a=a_tg^t$ for some $t=0,\ldots,k-1$, $a_t\in\F_p^*$, we have that either $\sigma_1(a)=0$ or $\sigma_2(a)=0$. Without loss of generality, we let   $\sigma_2(a)=0$. Then, denoting $x=\sigma_1(x),\alpha=\sigma_1(a)=a_tg_1^t$, we have
\begin{equation}
\sum_{z\in\F_p^n}\zeta_p^{\Tr_m(b(H(z+a)-cH(z)))}=p^{n_2}\sum_{x\in\F_p^{n_1}}\zeta_p^{\Tr_m(b(F(x+\alpha)-cF(x))}.
\end{equation}
 The autocorrelation of $H$ with respect to  $a$ is zero if and only if the autocorrelation of $F$ with respect to  $\alpha$ is zero.

Taking into account the two cases, $\sigma_1(a)=0$ or $\sigma_2(a)=0$, the theorem follows.
\end{proof}
\begin{defn}
Function $F: \F_{p^n} \rightarrow \F_{p^n}$ is called $\frac{1}{p}$-$c$-independent in its $x$ input if and only if for any $c,b,x \in \F_{p^n}$, $\alpha\in \F_{p}$ and $a = a_{t}g^{t}, t =0,1\ldots,k-1,  a_{t}\in F_{p}^*$, the probability $Prob(Tr(bF(x+a)) = Tr(bcF(x))+\alpha) =\frac{1}{p}$.
\end{defn}

In order to connect it to the $c$-SAC condition, we can say that, if the function $F$ is $\frac{1}{p}$-$c$-independent in all components, then it is $c$-SAC, meaning that $\sum_{x\in\F_{p^n}}\zeta^{\Tr_m(bF(x+a))-\Tr_m(cbF(x))}=0$.

\begin{thm} 
If $\cW_F(x,b)\overline{\cW_{F}(-x,bc)}=\cW_F(x+z,b)\overline{\cW_{F}(-x-z,bc)}$ for any $z = \sum\limits_{i=0}^{n-1} a_{i} g^i$, so that $a \in \F_{p}$ and $g$, a generator of the field,  $ z\in I_{i_{1} i_{2} \ldots i_{m}}=\left\{a_{0}g^0+ \cdots+ a_{n-1}g^{n-1} \mid a_{i} \neq 0 \Longrightarrow i \in\left\{i_{0}, \ldots, i_{m-1}\right\}\right\}$, then $F(x)$ is $\frac{1}{p}$-$c$-independent in the input coordinates $a_{i_0},a_{i_1},\ldots,a_{i_n-1} $.
\end{thm}

\begin{proof}
Let $x' \in \F_{p^n}$, $x' = \sum\limits_{i=0}^{n-1}  a'_{i}g^i$, and 
$$ S_{x'} = \{x \in \F_{p^n} \mid x' = \sum\limits_{i=0}^{n-1} a'_{i}g^i = x'_{m}\},$$ so that, $$\F_{p^{n}}=\bigcup_{x^{\prime} \in G F(p)^{m}} S_{x^{\prime}}, \quad S_{x_{1}^{\prime}} \cap S_{x_{2}^{\prime}}=\emptyset \Longleftrightarrow x_{1}^{\prime} \neq x_{2}^{\prime}.$$

By the hypothesis of the theorem, we can write:
$$\sum_{x \in S_{x'}}{\cW_F(x,b)\overline{\cW_{F}(-x,bc)}}=\sum_{x \in S_{x'+z'}}{\cW_F(x,b)\overline{\cW_{F}(-x,bc)}},$$
for any $x', z' \in \F_{p^n}$. Now let $x' = a_0g^0 + x''$, so that  $x'' = \sum\limits_{i=1}^{n-1} a_{i} g^i$ and $a_0 = 0$, or for short notation $x' = (0,x'')$, and the same for $z' = (j,z''), z'' \in I_{i_{2} \ldots i_{m}},  1 \leq j \leq p -1$, will be
\begin{equation*}
\begin{split}
&\sum_{x \in S_{(0,x'')}}{\cW_F(x,b)\overline{\cW_{F}(-x,bc)}}\\
&= \sum_{x \in S_{(j,x''+z'')}}{\cW_F(x,b)\overline{\cW_{F}(-x,bc)}},
\end{split}
\end{equation*}
for any $x'', z'' \in \F(p^{n-1})$. Thus,

\begin{align*}
&\sum_{x'' \in F_{p^{n-1}}}\sum_{x \in S_{(0,x'')}}{\cW_F(x,b)\overline{\cW_{F}(-x,bc)}}\\
&\qquad =\sum_{x'' \in F_{p^{n-1}}}\sum_{x \in S_{(j,x''|z'')}}{\cW_F(x,b)\overline{\cW_{F}(-x,bc)}}
\end{align*}
and
\begin{equation}
\begin{split}
&\sum_{x:x_{i_1}=0}{\cW_F(x,b)\overline{\cW_{F}(-x,bc)}}\\
&\qquad =\sum_{x:x_{i_1}=j}{\cW_F(x,b)\overline{\cW_{F}(-x,bc)}}\\
& \text {for }j\in {1,2\ldots,p-1}.
\end{split}
\end{equation}
That means that $f(x)$ is $\frac{1}{p}$-$c$-independent in the $i_1$ input and the other components can be obtained in the same way.
\end{proof}

\section{\uppercase{The $c$-Strict Avalanche Criterion of higher order}}
\label{sec:The $c$-Strict Avalanche Criterion of higher order}

Given a function $F: \F_{p^n} \rightarrow \F_{p^l}$, we fix a set of indices $I = \{j_1, \ldots, j_m\}$, and define a restriction of $F$, by fixing the coordinates corresponding to the indices in $I$, that is,  the restriction's input is written as $x = \sum\limits_{i \notin I} a_i g^i + \sum\limits_{j \in I} a_j g^j $, where $a_j \in \F_p$ are chosen to be constants. Then we can define the {\em $c$-Strict Avalanche Criterion of order $m$ ($c$-SAC(m))} as follows.

\begin{defn} 
Let $F$ be an $(n,m)$-vectorial $p$-ary function. Then, the function $F(x)$ satisfies the {\em $c$-Strict Avalanche Criterion of order $m$ ($c$-SAC(m))} if for $m$ chosen constant inputs, the corresponding restriction of $F(x)$ satisfies $c$-SAC.
\end{defn}

\begin{thm}
If a function $F(x)$ satisfies $c$-SAC($m$), it also satisfies $c$-SAC ($m-1$), for $1\leq m\leq n-1$.
\end{thm}
\begin{proof}
First we introduce the function ${F_I}_{c_{j_{1}} \ldots c_{j_{m}}}^{z_{j_{1}} \cdots z_{j_{m}}}(x)$, which is obtained from the function $F(x)$ by fixing $a_j$ to a constant $y_j$. So the input of the $F_I$ will be of the form $x = \sum\limits_{i \notin I} a_i g^i + \sum\limits_{j \in I} a_j g^j $, where $I = \{j_1 \ldots {j_{m}}\}$ is a set of indices of fixed elements for $F_I(x)$.

Thus,  $I = \{j_1, \ldots, {j_{m}}\}$ is a set of indices of fixed elements for $F(x)$, where $x = \sum\limits_{i \notin I} a_i g^i + \sum\limits_{j \in I} a_j g^j $, and $z_j = a_j g^j =  y_j g^j$ where $y_j \in \F_p$ are fixed constants.

Then, $F(x)$ can be written as $F_{c_{j_{1}} \ldots c_{j_{m}}}^{z_{j_{1}} \cdots z_{j_{m}}}(x)$, obtained from $F(x)$ by fixing $a_j$ to constant $y_j$, and $\alpha_i = u_i g^i$, $u_i \in \F_p, i \notin I$. We next consider:
\begin{equation}
M = \sum\limits_{x'= \sum\limits_{i \notin I} a_i g^i, a_i \in F_p}\zeta^{\Tr(bF_{c_{j_{1}} \ldots c_{j_{m}}}^{z_{j_{1}} \cdots z_{j_{m}}}(x+\alpha))-\Tr(bcF_{c_{j_{1}} \ldots c_{j_{m}}}^{z_{j_{1}} \cdots z_{j_{m}}}(x))}.
\end{equation}
Now, let  $x = \sum\limits_{i \notin I} a_i g^i + \sum\limits_{j \in I \setminus \{j_l\}} a_j g^j + \delta g^{j_l}$, where $\delta $ is an element extracted from the set $I$ corresponding to $g^{j_l}$, so the function will be of the form $F_{c_{j_{1}} \ldots c_{j_{l-1}},c_{j_{l+1}} \ldots c_{j_{m}},\delta}^{z_{j_{1}} \ldots z_{j_{l-1}},z_{j_{l+1}} \ldots z_{j_m},z_{j_{l}}}$, which, for brevity, it will be denoted $F_{c_{j_{1}} \ldots c_{j_{m},}\delta}^{z_{j_{1}} \ldots z_{j_{m}} z_{j_l}}$. 
Next,
\begin{equation}
M = \sum\limits_{\delta=0}^{p-1}\sum_{\substack{x'= \sum\limits_{i \notin I} a_i g^i,\\a_i \in F_p}}\zeta^{\Tr(bF_{c_{j_{1}} \ldots c_{j_{m},}\delta}^{z_{j_{1}} \ldots z_{j_{m}} z_{j_l}} (x+\alpha))-\Tr(cbF_{c_{j_{1}} \ldots c_{j_{m},}\delta}^{z_{j_{1}} \ldots z_{j_{m}} z_{j_l}}(x))}.
\end{equation}

Following the definition of the $c$-SAC all internal $p$-sums are equal to zero, since the function satisfies $c$-SAC($m$). Therefore, for $M = 0$, the function satisfies the $c$-SAC($m-1$), since every $F_{c_{j_{1}} \ldots c_{j_{m}}}^{z_{j_{1}} \ldots z_{j_{m}}}$ satisfies $c$-SAC.
\end{proof}

\section{\uppercase{Computational results and analysis}}
\label{computations}

This section displays a partial list of functions of type  $F: \F_{2^3} \rightarrow \F_{2^3}$ and $F:  \F_{3^2} \rightarrow  \F_{3^2}$ that fulfill $c$-SAC but are not PcN (which, for $n=m$, is equivalent to $c$-bent$_1$), for the values of $c$ given in the list. It is interesting to note that all functions of type $F: \F_{2^2} \rightarrow \F_{2^2}$ we found that fulfilled $c$-SAC were  PcN for those values of $c$.

As argued before, for $p=2$ there are no $(n,n)$-functions that are $c$-SAC for $c=1$. We do, however, find examples of $(n,n)$-functions that fulfill $c$-SAC for $c=1$ for $p=3$. These are, however, PN (perfect nonlinear). Below, $g$ denotes a primitive element in the considered field.

\subsection{Even characteristic}

The following functions  $F: \F_{2^3} \rightarrow \F_{2^3}$ fulfill $c$-SAC for $c \in \{ g, g^2, g^2 + g + 1, g^2 + 1\}$: 
\begin{enumerate}
\item $(g^2 + 1)x^6 + (g + 1)x^5 + (g + 1)x^4 + (g^2 + g)x^3 + x^2 + g^2x$;
\item $gx^6 + x^5 + gx^4 + (g^2 + g + 1)x^3 + (g + 1)x^2 + (g + 1)x$;
\item $g^2x^6 + (g^2 + g + 1)x^5 + x^4 + (g^2 + g)x^3 + (g^2 + g)x^2 + (g^2 + 1)x$;
\item $(g^2 + g + 1)x^6 + gx^5 + x^4 + x^3 + (g^2 + g)x^2 + (g^2 + 1)x$;
\item $(g^2 + g + 1)x^6 + (g^2 + g)x^5 + g^2x^3 + (g^2 + 1)x^2 + (g^2 + g + 1)x$;
\item $(g^2 + g)x^6 + (g + 1)x^5 + g^2x^4 + gx^3 + gx^2 + g^2x$;

\item $g^2x^6 + (g^2 + 1)x^5 + gx^4 + x^3 + g^2x^2$;
\item $(g^2 + g)x^6 + (g + 1)x^5 + (g^2 + 1)x^4 + gx^3 + (g + 1)x^2 + (g^2 + g + 1)x$;
\item $g^2x^6 + (g^2 + 1)x^5 + (g^2 + 1)x^4 + x^3 + (g + 1)x^2 + (g^2 + g + 1)x$;
\item $gx^6 + (g^2 + g)x^5 + (g + 1)x^4 + (g + 1)x^3 + x^2 + g^2x$;
\item $(g^2 + g + 1)x^6 + gx^5 + (g + 1)x^4 + x^3 + x^2 + g^2x$;
\item $(g^2 + g)x^6 + (g + 1)x^5 + (g^2 + g)x^4 + gx^3 + (g^2 + 1)x^2 + (g^2 + 1)x$;
\item $gx^6 + x^5 + (g^2 + g + 1)x^3 + g^2x^2 + gx$;
\item $x^6 + (g^2 + 1)x^5 + (g^2 + g + 1)x^4 + (g^2 + g)x^3 + (g^2 + g)x^2 + (g^2 + g + 1)x$;
\item $(g^2 + 1)x^6 + (g + 1)x^5 + (g^2 + 1)x^4 + (g^2 + g)x^3 + gx^2 + x$;
\item $(g^2 + g + 1)x^6 + gx^5 + (g^2 + g)x^4 + x^3 + (g^2 + g + 1)x^2 + (g^2 + g + 1)x$;
\item $gx^6 + x^5 + x^4 + (g^2 + g + 1)x^3 + (g^2 + g)x^2 + (g^2 + 1)x$;
\item $(g^2 + 1)x^6 + (g + 1)x^5 + x^4 + (g^2 + g)x^3 + (g^2 + g)x^2 + (g^2 + 1)x$;
\item $(g^2 + g + 1)x^6 + g^2x^5 + (g^2 + g + 1)x^4 + (g + 1)x^3 + x^2 + (g^2 + 1)x$;
\item $x^6 + (g^2 + 1)x^5 + (g^2 + 1)x^4 + (g^2 + g)x^3 + (g + 1)x^2 + (g^2 + g + 1)x$;
\item $(g^2 + g)x^6 + x^5 + (g + 1)x^4 + (g^2 + 1)x^3 + (g^2 + g)x^2 + (g^2 + g)x$;
\item $(g^2 + g + 1)x^6 + g^2x^5 + (g^2 + 1)x^4 + (g + 1)x^3 + (g + 1)x^2 + (g^2 + g + 1)x$;
\item $gx^6 + (g^2 + g + 1)x^5 + x^4 + (g^2 + 1)x^3 + gx^2$;
\item $(g + 1)x^6 + g^2x^5 + (g^2 + g + 1)x^4 + x^3 + g^2x^2 + (g^2 + g)x$;
\item $x^6 + (g + 1)x^5 + (g^2 + g)x^4 + g^2x^3 + (g^2 + 1)x$;
\item $(g^2 + g)x^6 + gx^5 + (g^2 + g)x^4 + g^2x^3 + (g^2 + 1)x^2 + (g^2 + g + 1)x$;
\item $(g^2 + 1)x^6 + (g^2 + g + 1)x^5 + (g^2 + g + 1)x^4 + (g + 1)x^3 + g^2x^2 + (g^2 + g)x$.
\end{enumerate}
The following functions  $F: \F_{2^3} \rightarrow \F_{2^3}$ fulfill $c$-SAC for $c  \in \{ g, g + 1, g^2 + g, g^2 + 1\}$:
\begin{enumerate}
\item $(g^2 + g)x^6 + gx^5 + (g^2 + 1)x^4 + g^2x^3 + g^2x^2 + (g + 1)x$; 
\item $(g^2 + 1)x^6 + (g^2 + g + 1)x^5 + g^2x^4 + (g + 1)x^3 + (g^2 + 1)x^2 + gx$;
\item $(g + 1)x^6 + (g^2 + 1)x^5 + (g^2 + g + 1)x^4 + (g^2 + g + 1)x^3 + gx^2 + (g^2 + g + 1)x$;
\item $(g^2 + g)x^6 + gx^5 + g^2x^4 + g^2x^3 + (g^2 + 1)x^2 + gx$; \item $(g + 1)x^6 + (g^2 + 1)x^5 + x^4 + (g^2 + g + 1)x^3 + (g^2 + g)x^2 + (g^2 + 1)x$;
\item $gx^6 + (g^2 + g)x^5 + x^4 + (g + 1)x^3 + (g^2 + g)x^2 + (g^2 + 1)x$;
\item $(g^2 + 1)x^6 + (g^2 + g + 1)x^5 + (g + 1)x^4 + (g + 1)x^3 + x$.
\item $x^6 + (g^2 + g)x^5 + g^2x^4 + (g^2 + g + 1)x^3 + (g^2 + g + 1)x^2 + (g^2 + 1)x$;
\item $g^2x^6 + gx^5 + (g^2 + g + 1)x^4 + (g^2 + 1)x^3 + g^2x^2 + (g^2 + g)x$;
\item $(g^2 + 1)x^6 + g^2x^5 + gx^4 + gx^3 + (g^2 + g)x^2 + gx$;
\item $x^6 + (g^2 + g)x^5 + (g^2 + g + 1)x^4 + (g^2 + g + 1)x^3 + g^2x^2 + (g^2 + g)x$;
\item $(g^2 + 1)x^6 + g^2x^5 + (g + 1)x^4 + gx^3 + x^2 + g^2x$;
\item $(g^2 + g)x^6 + x^5 + (g + 1)x^4 + (g^2 + 1)x^3 + x^2 + g^2x$;
\item $g^2x^6 + gx^5 + (g^2 + 1)x^4 + (g^2 + 1)x^3 + (g + 1)x$;
\item $(g + 1)x^6 + x^5 + x^4 + gx^3 + (g^2 + g + 1)x$;
\item $(g^2 + g + 1)x^6 + g^2x^5 + (g + 1)x^3 + (g^2 + 1)x^2 + (g + 1)x$;
\item $(g^2 + g)x^6 + x^5 + g^2x^4 + (g^2 + 1)x^3 + gx^2$;
\item $x^6 + (g^2 + 1)x^5 + gx^4 + (g^2 + g)x^3 + (g^2 + g + 1)x^2 + x$;
\item $(g^2 + g + 1)x^6 + g^2x^5 + gx^4 + (g + 1)x^3 + (g^2 + g + 1)x^2 + x$;
\item $(g^2 + 1)x^6 + (g + 1)x^5 + gx^4 + (g^2 + g)x^3 + (g^2 + g)x^2 + (g^2 + g + 1)x$;
\item $gx^6 + x^5 + (g^2 + g)x^4 + (g^2 + g + 1)x^3 + gx^2 + (g + 1)x$;
\item $(g^2 + g + 1)x^6 + gx^5 + x^4 + x^3 + (g + 1)x^2 + x$;
\item $(g^2 + 1)x^6 + (g + 1)x^5 + (g^2 + g)x^4 + (g^2 + g)x^3 + gx^2 + (g + 1)x$;
\item $(g^2 + 1)x^6 + (g^2 + g + 1)x^5 + (g^2 + g)x^4 + (g + 1)x^3 + (g + 1)x^2 + (g^2 + g)x$;
\item $gx^6 + x^5 + (g^2 + g + 1)x^4 + (g^2 + g + 1)x^3 + g^2x$;
\item $(g^2 + g)x^6 + gx^5 + g^2x^3 + (g^2 + g + 1)x^2 + g^2x$;
\item $(g + 1)x^6 + (g^2 + 1)x^5 + gx^4 + (g^2 + g + 1)x^3 + x^2$;
\item $(g^2 + 1)x^6 + (g^2 + g + 1)x^5 + x^4 + (g + 1)x^3 + (g^2 + g)x^2 + (g^2 + 1)x$.
\end{enumerate}
The following functions  $F: \F_{2^3} \rightarrow \F_{2^3}$ fulfill $c$-SAC for 
$c  \in \{ g^2, g + 1, g^2 + g, g^2 + g + 1\}$ :
\begin{enumerate}
\item $gx^6 + (g^2 + g + 1)x^5 + g^2x^4 + (g^2 + 1)x^3 + x^2 + (g^2 + g + 1)x$;
\item $(g + 1)x^6 + g^2x^5 + gx^4 + x^3 + (g^2 + g + 1)x^2 + x$;
\item $x^6 + (g + 1)x^5 + (g + 1)x^4 + g^2x^3 + (g + 1)x^2 + gx$;
\item $gx^6 + (g^2 + g + 1)x^5 + gx^4 + (g^2 + 1)x^3 + (g^2 + g + 1)x^2 + x$;
\item  $x^6 + (g + 1)x^5 + x^4 + g^2x^3 + (g^2 + g)x^2 + (g^2 + 1)x$;
\item $(g^2 + 1)x^6 + g^2x^5 + x^4 + gx^3 + (g^2 + g)x^2 + (g^2 + 1)x$;
\item  $(g + 1)x^6 + g^2x^5 + (g^2 + g)x^4 + x^3 + g^2x^2$;
\item $g^2x^6 + gx^5 + (g^2 + g + 1)x^4 + (g^2 + 1)x^3 + (g^2 + g + 1)x^2 + gx$;
\item $(g^2 + g)x^6 + gx^5 + (g^2 + g + 1)x^4 + g^2x^3 + (g + 1)x^2 + gx$;
\item $(g^2 + 1)x^6 + (g^2 + g + 1)x^5 + (g^2 + g)x^4 + (g + 1)x^3 + gx^2 + (g + 1)x$;
\item $(g + 1)x^6 + (g^2 + 1)x^5 + (g^2 + 1)x^4 + (g^2 + g + 1)x^3 + (g^2 + 1)x^2 + (g^2 + g)x$;
\item $(g^2 + g)x^6 + gx^5 + (g^2 + g)x^4 + g^2x^3 + gx^2 + (g + 1)x$;
\item $(g + 1)x^6 + (g^2 + 1)x^5 + (g + 1)x^4 + (g^2 + g + 1)x^3 + x^2 + g^2x$;
\item $g^2x^6 + (g^2 + g + 1)x^5 + (g + 1)x^4 + (g^2 + g)x^3 + x^2 + g^2x$;
\item $(g^2 + 1)x^6 + (g^2 + g + 1)x^5 + x^4 + (g + 1)x^3 + (g^2 + g + 1)x^2$;
\item $(g^2 + g + 1)x^6 + (g^2 + g)x^5 + gx^4 + g^2x^3 + gx^2 + (g^2 + g)x$;
\item $g^2x^6 + (g^2 + g + 1)x^5 + (g^2 + g)x^3 + (g + 1)x^2 + (g^2 + 1)x$;
\item $(g^2 + g + 1)x^6 + (g^2 + g)x^5 + g^2x^4 + g^2x^3 + (g^2 + 1)x^2 + gx$;
\item $(g + 1)x^6 + x^5 + g^2x^4 + gx^3 + (g^2 + 1)x^2 + gx$;
\item $(g + 1)x^6 + x^5 + (g^2 + g)x^4 + gx^3 + g^2x^2 + x$;
\item $(g^2 + g + 1)x^6 + (g^2 + g)x^5 + (g + 1)x^4 + g^2x^3 + x^2 + g^2x$;
\item $g^2x^6 + (g^2 + g + 1)x^5 + (g^2 + g + 1)x^4 + (g^2 + g)x^3 + (g^2 + g + 1)x^2 + (g + 1)x$;
\item $(g + 1)x^6 + x^5 + (g + 1)x^4 + gx^3 + x^2 + g^2x$;
\item $g^2x^6 + (g^2 + g + 1)x^5 + g^2x^4 + (g^2 + g)x^3 + (g^2 + 1)x^2 + gx$;
\item $gx^6 + (g^2 + g)x^5 + g^2x^4 + (g + 1)x^3 + (g^2 + 1)x^2 + gx$;
\item $g^2x^6 + (g^2 + 1)x^5 + (g^2 + 1)x^4 + x^3 + (g^2 + g)x$;
\item $gx^6 + (g^2 + g)x^5 + (g + 1)x^3 + g^2x^2 + (g^2 + g + 1)x$;
\item $(g^2 + g)x^6 + (g + 1)x^5 + gx^4 + gx^3 + (g^2 + g + 1)x^2 + x$;
\item $g^2x^6 + (g^2 + 1)x^5 + gx^4 + x^3 + (g^2 + g + 1)x^2 + x$.

\end{enumerate}

\subsection{Odd characteristic}
The following functions  $F:  \F_{3^2} \rightarrow  \F_{3^2}$    all fulfill $c$-SAC and are PcN for $c = 1$, in addition to the values of $c$ displayed.

For   $c \in \{g, 2g, 2g + 2\}$:
\begin{enumerate}
\item $(g + 2)x^6 + gx^4 + x^3 + (2g + 1)x^2 + 2gx$;
\item $2x^6 + 2x^4 + (g + 1)x^3 + 2x^2 + 2gx$;
\item $(2g + 2)x^6 + (2g + 2)x^4 + x^3 + (2g + 2)x^2 + gx $;
\item $(g + 1)x^6 + x^4 + gx^3 + (2g + 2)x^2 + x$;
\item $gx^6 + (2g + 1)x^4 + (2g + 1)x^3 + 2gx^2 + 2x$;
\item $(g + 1)x^6 + (g + 1)x^4 + (g + 1)x^3 + (g + 1)x^2$;
\item $gx^6 + gx^4 + gx^2 + 2gx$;
\item $gx^6 + (2g + 1)x^4 + 2x^3 + 2gx^2$;
\item $(g + 1)x^6 + x^4 + (g + 2)x^3 + (2g + 2)x^2$;
\item $x^6 + (2g + 2)x^4 + 2x^3 + 2x^2 + gx$;
\item $(g + 2)x^6 + 2gx^4 + gx^3 + (2g + 1)x^2 + 2x$;
\item $2x^6 + 2x^4 + gx^3 + 2x^2 + (2g + 2)x$;
\item $gx^6 + gx^4 + gx^3 + gx^2$;
\item $(g + 1)x^6 + x^4 + (2g + 2)x^2 + gx$;
\item $gx^6 + gx^4 + x^3 + gx^2 + (g + 2)x$;
\item $(g + 2)x^6 + 2gx^4 + (g + 1)x^3 + (2g + 1)x^2 + (2g + 1)x$;
\item $2gx^6 + (g + 2)x^4 + gx^2 + (2g + 2)x$;
\item $2x^6 + 2x^4 + (2g + 2)x^3 + 2x^2 + (g + 2)x$;
\item $x^6 + (2g + 2)x^4 + (g + 2)x^3 + 2x^2 + (2g + 2)x$;
\item $2x^6 + (2g + 2)x^4 + x^2 + 2gx$;
\item $(2g + 1)x^6 + (2g + 1)x^4 + 2gx^3 + (2g + 1)x^2 + (2g + 2)x$;
\item $(g + 2)x^6 + 2gx^4 + (2g + 2)x^3 + (2g + 1)x^2 + 2gx$;
\item $2x^6 + (2g + 2)x^4 + 2gx^3 + x^2 + (g + 1)x$;
\item $gx^6 + gx^4 + (g + 2)x^3 + gx^2 + (g + 1)x$;
\item $2x^6 + (2g + 2)x^4 + gx^3 + x^2 + 2x$;
\item $gx^6 + (g + 2)x^4 + 2gx^2 + x$;
\item $x^6 + x^4 + (g + 2)x^3 + x^2 + (2g + 2)x $;
\item $x^6 + (g + 1)x^4 + (g + 1)x^3 + 2x^2 + (2g + 1)x$;
\item $x^6 + x^4 + x^2 + 2x$;
\item $x^6 + (g + 1)x^4 + (2g + 1)x^3 + 2x^2$;
\item $gx^6 + (g + 2)x^4 + (2g + 2)x^3 + 2gx^2$;
\item $(g + 2)x^6 + 2gx^4 + (2g + 1)x^3 + (2g + 1)x^2 + x$;
\item $(2g + 2)x^6 + 2x^4 + x^3 + (g + 1)x^2 + (2g + 1)x$;
\item $(2g + 2)x^6 + 2x^4 + gx^3 + (g + 1)x^2 + (g + 1)x$;
\item $(2g + 1)x^6 + (2g + 1)x^4 + x^3 + (2g + 1)x^2 + 2gx$.

\end{enumerate}

For   $c\in \{ g + 1, 2g + 1, g + 2\}$: 
\begin{enumerate}
\item $(2g + 1)x^6 + 2gx^4 + 2x^3 + (g + 2)x^2 + (g + 2)x$;
\item $ x^6 + x^4 + (2g + 2)x^3 + x^2 + (g + 2)x$;
\item $2gx^6 + (2g + 1)x^4 + 2gx^3 + gx^2 + (g + 1)x$;
\item $(2g + 1)x^6 + 2gx^4 + x^3 + (g + 2)x^2 + 2gx$;
\item $(2g + 2)x^6 + 2x^4 + 2x^3 + (g + 1)x^2 + (g + 2)x$;
\item $2gx^6 + (g + 2)x^4 + 2x^3 + gx^2$;
\item $(2g + 2)x^6 + (2g + 2)x^4 + (g + 2)x^3 + (2g + 2)x^2 + 2x$;
\item $2gx^6 + 2gx^4 + (g + 2)x^3 + 2gx^2 + (g + 1)x$;
\item $2x^6 + (g + 1)x^4 + 2gx^3 + x^2$;
\item $2x^6 + (2g + 2)x^4 + x^3 + x^2 + (g + 2)x$;
\item $2gx^6 + (g + 2)x^4 + gx^3 + gx^2 + x$;
\item $(2g + 2)x^6 + 2x^4 + 2gx^3 + (g + 1)x^2 + (2g + 2)x$;
\item $2x^6 + (2g + 2)x^4 + (2g + 1)x^3 + x^2$;
\item $2gx^6 + (2g + 1)x^4 + (2g + 1)x^3 + gx^2 + (2g + 2)x $;
\item $(g + 1)x^6 + x^4 + 2gx^3 + (2g + 2)x^2 + 2x$;
\item $x^6 + (g + 1)x^4 + 2x^2 + 2gx$;
\item $2gx^6 + 2gx^4 + gx^3 + 2gx^2$;
\item $2x^6 + (g + 1)x^4 + x^2 + (g + 2)x$;
\item $(2g + 1)x^6 + gx^4 + (2g + 2)x^3 + (g + 2)x^2 + 2gx$;
\item $2gx^6 + (g + 2)x^4 + (2g + 1)x^3 + gx^2 + 2x$;
\item $gx^6 + (2g + 1)x^4 + gx^3 + 2gx^2 + x$;
\item $x^6 + x^4 + (2g + 1)x^3 + x^2 + (g + 1)x$;
\item $gx^6 + (2g + 1)x^4 + (2g + 2)x^3 + 2gx^2 + (2g + 1)x$;
\item $2x^6 + 2x^4 + (2g + 2)x^3 + 2x^2 + gx$;
\item $x^6 + x^4 + x^2 + x$;
\item $2gx^6 + 2gx^4 + (2g + 2)x^3 + 2gx^2 + (2g + 1)x$;
\item $x^6 + (g + 1)x^4 + 2gx^3 + 2x^2 + (g + 1)x$;
\item $(g + 2)x^6 + (g + 2)x^4 + (g + 2)x^2 + (g + 2)x$;
\item $x^6 + (g + 1)x^4 + gx^3 + 2x^2 + 2x$;
\item $(g + 1)x^6 + x^4 + (g + 1)x^3 + (2g + 2)x^2 + (g + 2)x$;
\item $2gx^6 + 2gx^4 + (2g + 1)x^3 + 2gx^2 + 2x$;
\item $2x^6 + 2x^4 + x^3 + 2x^2$;
\item $(2g + 1)x^6 + gx^4 + 2x^3 + (g + 2)x^2$;
\item $2gx^6 + (2g + 1)x^4 + 2x^3 + gx^2 + 2gx$;
\item $ (g + 2)x^6 + gx^4 + 2x^3 + (2g + 1)x^2 + gx$;
\end{enumerate}

\section{\uppercase{Conclusion and future work}}
\label{sec:Conclusion and future work}

In this paper, we have generalized the concept of Strict Avalanche Criterion (SAC) to address possible $c$-differential attacks, in the realm of finite fields. Further, we have defined the concepts of $c$-Strict Avalanche Criterion ($c$-SAC) and $c$-Strict Avalanche Criterion of order $m$ ($c$-SAC($m$)), and generalized results of \cite{LC07}. By computing and checking functions of the given type, we have also shown that the new definition is not equivalent to the existing concepts of $c$-bent$_1$-ness \cite{SGGRT20}, nor (for $n=m$) PcN-ness \cite{EFRST20}. It would of interest,  to find, theoretically, classes of functions that fulfill $c$-SAC or $c$-SAC($m$) for large $n$ and $m$, and to find other properties satisfied by $c$-SAC functions, as well as devise a practical attack on particular S-boxes using these concepts. Finally, for small examples, all functions that we found that fulfilled 1-SAC for $n=m$ were PN. It would be interesting to find either a function which fulfills 1-SAC but is not PN, or a proof that this cannot happen.

\bibliographystyle{apalike}
{\small
\bibliography{cSAC}}

\end{document}